\documentclass[prl, showpacs, twocolumn, floatfix]{revtex4}

\usepackage{graphicx}
\usepackage{amsmath, amsfonts, amssymb, bm}
\usepackage[]{psfrag}

\psfragscanoff
\begin{document} 

\title{ Resonant two-photon single ionization of two atoms   }

\author{ C. M{\"u}ller and A. B. Voitkiv }  
\affiliation{Max-Planck-Institut f\"ur Kernphysik, Saupfercheckweg 1, 69117 Heidelberg, Germany}

\date{\today}

\begin{abstract}

Resonant two-photon ionization in a system consisting 
of two spatially well-separated atoms is studied. 
Due to two-center electron-electron correlations, 
the ionization may also proceed through
photo-excitation of both atoms with subsequent interatomic Coulombic
decay. We show that this channel may dominate the photoionization process 
and qualitatively change its dependence on the field intensity 
and the spectra of emitted electrons. 

\end{abstract}
 
\pacs{
32.80.Rm, 
32.80.Zb,  
42.50.Hz 
} 

\maketitle 

The development of tunable laser sources in the 1970s stimulated
in-depth studies of the behavior of atoms and molecules exposed to
intense resonant electromagnetic fields \cite{knight-review,fedorov}. 
While the emission of fluorescence light can provide detailed atomic structure
information, resonance-enhanced multiphoton ionization has become a
valuable tool for probing dynamical aspects of atoms,
clusters, biomolecules and chemical reactions \cite{REMPI}.

Resonant two-photon ionization of atomic hydrogen has been examined
thoroughly by theoreticians as a prime example which allows a detailed
understanding of the underlying field-induced dynamics. Due to the
resonantly driven bound-bound transition, the photoelectron spectra 
display Autler-Townes doublets and the total ionization probability
exhibits a step-wise temporal evolution \cite{knight-review}. 
Other nonperturbative effects in strong laser fields 
were studied as well (see e.g. \cite{Rzazewski,Dieter} 
and references therein). Corresponding experimental studies 
are becoming feasible nowadays due to the advent 
of brilliant high-frequency photon sources such as
advanced synchrotron beam lines and free-electron lasers. 

Ionization may also occur through resonant photoexcitation of an
autoionizing atomic level, which subsequently decays via Auger emission.
In recent years this type of ionization mechanism, which relies on
electron-electron correlations, has been investigated intensively, both
theoretically \cite{ICD,ICDtheo,Najjari} and experimentally, in systems consisting 
of two (or more) atoms, such as noble gas dimers \cite{dimers,helium-dimers}, metal
oxides \cite{MARPE} or water molecules \cite{water}. Here the
radiationless deexcitation of one atom is accompanied with the 
ionization of another neighboring atom. This process is commonly
referred to as interatomic Coulombic decay, ICD \cite{ICD}. 

\begin{figure}[b]  
\vspace{-0.25cm}
\begin{center}
\includegraphics[width=0.33\textwidth]{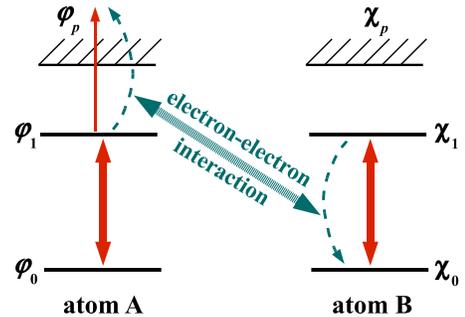}
\end{center}
\vspace{-0.5cm} 
\caption{ \footnotesize{ Scheme of 
resonant two-photon single ionization of two atoms. 
In order not to overload the picture,   
ionization channels are shown for atom $A$ only. 
Ionization of atom $B$ proceeds analogously. }} 
\label{figure1}
\end{figure}

In this Letter, we study resonant two-photon ionization of a
two-atom system which is exposed to an external electromagnetic field.
In this situation, each atom is subject not only to the influence of the
field but also to the interaction with the neighbor atom. Hence there
are two pathways for ionization: either directly via resonant two-photon absorption 
at a single center, or mediated by ICD when each atom has been excited 
by single-photon absorption (see figure \ref{figure1}). 
We shall show that the second mechanism 
can be remarkably efficient and may even dominate the direct channel by orders 
of magnitude. Moreover, the two-center channel exhibits very peculiar features 
with respect to its field dependence, temporal development and photoelectron spectra.  
In contrast to the two-center resonant photoionization considered in \cite{Najjari}, 
the present process occurs for two identical atoms. It can, in principle, 
also exist in the case of two different atomic species 
provided that the latter ones possess a common dipole-allowed 
transition frequency.  

Let us consider a system consisting of two atoms 
($A$ and $B$) separated by a sufficiently large 
distance $R$ such that their individuality 
is basically preserved. 
Let each atom have an excited state 
reachable from its ground state 
by a dipole-allowed transition and 
the energy between these states 
be the same for both atoms. 
The atoms, which are initially in their ground states,  
are embedded in a resonant electromagnetic field.  
As seen from figure \ref{figure1}, 
already in the absence of the neighbor, 
each of the atoms can be ionized 
by absorbing two photons. However, due to ICD,  
photo ionization in the system of two atoms  
acquires interesting and 
qualitatively new features.  
   
Assuming the atoms to be at rest,  
we take the position of the nucleus of atom $A$  
as the origin and denote the coordinates 
of the nucleus of atom $B$, the electron of atom $A$ 
and that of atom $B$ by ${\bf R}$, ${\bf r}_1$ 
and ${\bf r}_2 ={\bf R} + \mbox{\boldmath{ $\xi$ }} $, 
respectively, where $\mbox{\boldmath{ $\xi$ }}$ 
is the position of the electron of atom $B$ 
with respect to its nucleus. 

The total Hamiltonian describing 
two atoms in the external electromagnetic field reads 
\begin{eqnarray} 
H =  \hat{H}_0 + \hat{V}_{AB} + \hat{W},  
\label{hamiltonian}  
\end{eqnarray} 
where $ \hat{H}_0 $ is the sum of the Hamiltonians 
for the noninteracting atoms $A$ and $B$,  
$\hat{V}_{AB}$ the interaction between the atoms 
and $\hat{W} = \hat{W}_A + \hat{W}_B $ the interaction 
of the atoms with the electromagnetic field.

For electrons undergoing  
electric dipole transitions the interatomic 
interaction $V_{AB}$ reads 
\begin{eqnarray} 
\hat{V}_{AB} &=& \frac{\alpha}{R^3} 
\left( e^{i k R} - i k R e^{-i k R} \right)   
- \frac{k^2 \beta}{R} e^{i k R}.  
\label{AB_inter} 
\end{eqnarray} 
Here, $\alpha = r_{1 i} \xi_j ( \delta_{ij} - 3 R_i R_j/R^2)$, 
$\beta = r_{1 i} \xi_j ( \delta_{ij} - R_i R_j/R^2)$,
$ r_{1 i} $ and $ \xi_j $ ($i,j \in \{ x,y,z \}$) 
are the coordinates of the electrons and  
a summation over the repeated indices is implied.  
Further, $k = \omega_{fi}/c$ where $\omega_{fi}$ 
is the transition frequency and $c$ the speed of light.  
Note that atomic units (a.u.) are used throughout 
unless otherwise stated.
 
The electromagnetic field will be treated  
as a classical, linearly polarized field,  
described by the vector potential 
${\bf A}({\bf r},t)= c {\bf F}_0/\omega_0 \cos\left(\omega_0 t - {\bf k}_0 \cdot {\bf r}\right)$,
where $\omega_0 = c k_0 $ and ${\bf k}_0$ 
are the angular frequency and wave vector,  
and ${\bf F}_0$ is the field strength. 
The interaction $\hat{W}$ then reads   
\begin{eqnarray} 
\hat{W} = \sum_{j=1,2} \frac{1}{c} {\bf A}({\bf r}_j,t) \cdot \hat{\bf p}_j,   
\label{interaction} 
\end{eqnarray} 
where $\hat{\bf p}_j$ is the momentum operator for 
the $j$-th electron. 
 
In what follows we shall assume that the electromagnetic field  
is not too strong such that field-induced transitions 
from bound states to the continuum are weak and, 
besides, one can neglect the influence of the field on 
the continuum. This also implies that 
only one of the centers, either $A$ or $B$, 
will be ionized with a non-negligible probability. 

However, even a relatively weak, but resonant, 
electromagnetic field can very effectively 
couple bound states belonging to the same center. 
Therefore, for each center we first construct 
field-dressed bound states. 
Taking, as an example, the center $A$,   
assuming that the field is switched on 
suddenly at $t_i = 0$, and using 
the rotating-wave approximation we obtain (see also \cite{knight-review,fedorov}) 
\begin{eqnarray}
\varphi_{+}(t) &=& \frac{1}{z_{+}^A - z_{-}^A}
\left( \left( z_{+}^A + \omega_0 - \varepsilon_1 \right) 
\exp(-i z_{+}^A t) \right.  
\nonumber \\ 
&-& \left.   \left( z_{-}^A + \omega_0 - \varepsilon_1 \right) 
\exp(-i z_{-}^A t) \right) \varphi_0 
\nonumber \\ 
&+& \frac{W_{10}}{z_{+}^A - z_{-}^A}
\left( \exp(-i z_{+}^A t) - \exp(-i z_{-}^A t) \right) 
\nonumber \\ 
&\times& \exp(- i \omega_0 t) \varphi_1 
\label{dressed-states-1} 
\end{eqnarray} 
and 
\begin{eqnarray}
\varphi_{-}(t) &=& \frac{ W_{01} }{ z_{+}^A - z_{-}^A }
\left( \exp(-i z_{+}^A t) - \exp(-i z_{-}^A t) \right) \varphi_0  
\nonumber \\ 
&+& \frac{1}{ z_{+}^A - z_{-}^A }
\left( \left( z_{+}^A - \varepsilon_0 \right) \exp(-i z_{+}^A t) 
\right.  
\nonumber \\ 
&-& \left. \left( z_{-}^A - \varepsilon_0 \right) 
\exp(-i z_{-}^A t) \right) \exp(- i \omega_0 t) \varphi_1.     
\label{dressed-states-2} 
\end{eqnarray} 
In these expressions $\varphi_0$ with an energy 
$\varepsilon_0$ and $\varphi_1$ with an energy   
$\varepsilon_1$ are the ground and excited states, 
respectively, of center $A$. Further, 
\begin{eqnarray}
z_{\pm}^A &=& \frac{1}{2} %
\left( \varepsilon_0 + \varepsilon_1 - \omega_0 \pm \Omega_R \right), %
\label{dressed-states-3}   
\end{eqnarray} 
where $\Omega_R = \sqrt{(\varepsilon_1 - \varepsilon_0 - \omega_0)^2 + 4 \left| W_{01} \right|^2} $ 
is the Rabi frequency 
and $W_{ij} = \left\langle \varphi_i \left| %
{\bf F}_0 \cdot \hat{\bf p}_1/(2 \omega_0) \right| \varphi_j \right\rangle $ 
($i,j \in \{0,1\}$). 
The states (\ref{dressed-states-1}) and (\ref{dressed-states-2}) 
are orthogonal to each other and to the continuum states 
$ \{ \varphi_{\bf p} \}$ of center $A$, and are normalized to unity.  
At $t=0$ these field-dressed states 
reduce to $\varphi_0$ and $\varphi_1$, respectively.  
In the above description we have neglected the spontaneous 
radiative decay of the excited state $\varphi_1$ which 
in our case is justified as long as $\left| W_{01} \right| \gg \Gamma_r$, 
where $\Gamma_r$ is the radiative width of $\varphi_1$. 

Expressions for the field-dressed bound states 
$\chi_{\pm}(t)$ on the center $B$ with the corresponding 
quasi-energies $z^B_{\pm}$ can be obtained from 
(\ref{dressed-states-1})-(\ref{dressed-states-3}) 
by straightforward replacements. 
  
Our consideration of the photoionization process  
will be based on the $S$-matrix formalism. Concentrating 
for the moment on the description of 
ionization of atom $A$ we write down the transition 
matrix element   
\begin{eqnarray} 
S_{fi} = - i \int_{t_i}^{t_f} dt \left\langle \psi^A_f(t) \left|  
\hat{V}_{AB} + \hat{W}_A  \right| \Psi(t) \right\rangle.   
\label{s-matrix}  
\end{eqnarray} 
Here $ \Psi $ is a solution of the Schr\"odinger equation  
for the total Hamiltonian 
$ H $ and $\psi^A_f $ is a final channel  
corresponding to ionization of $A$. Note that 
neither $\hat{W}_A$ nor $\hat{V}_{AB}$ are included 
in the Schr\"odinger equation for $\psi^A_f $. 

Taking into account that (i) the electromagnetic field-induced 
transitions to the continuum are weak, 
(ii) the field does not distort continuum states,  
(iii) the interatomic interaction 
at large distances is weak, and (iv) 
initially (at $t_i=0$) both atoms are 
in the ground states, one can approximate $\Psi(t)$ by 
$\varphi_{+}(t) \chi_{+}(t)$.  
Besides, the final state 
$\psi^A_f(t)$ can be represented by either 
one of the states $\varphi_{\bf p} \chi_{+}(t)$ and 
$\varphi_{\bf p} \chi_{-}(t)$. Here 
$ \varphi_{\bf p}$ is the continuum state 
of center $A$ with an asymptotic 
momentum ${\bf p}$ and 
energy $\varepsilon_p = p^2/2$.  

As a result, ionization of center $A$ 
is described by the following transition amplitudes      
\begin{eqnarray} 
S_{{\bf p},+}(t) &=& - i \int_0^{t} dt'   
\left\langle \varphi_{\bf p} \chi_{+} %
\left| \hat{ V }_{AB} + \hat{ W }_A \right| \varphi_{+} \chi_{+} \right\rangle  
\nonumber \\ 
S_{{\bf p},-}(t) &=& - i \int_0^{t} dt' 
\left\langle  \varphi_{\bf p} \chi_{-} %
\left| \hat{ V }_{AB} \right| \varphi_{+} \chi_{+} \right\rangle.   
\label{ioniz_of_center_A-add} 
\end{eqnarray} 
Since the final states 
$\varphi_{\bf p} \chi_{+}(t)$ and 
$\varphi_{\bf p} \chi_{-}(t)$ are 
orthogonal to each other, these amplitudes 
add up incoherently and for the probability of ionization 
of center $A$ we obtain 
\begin{eqnarray} 
P_A(t) &=& \int d^3{\bf p} 
\left( \mid  S_{{\bf p},+}(t) \mid^2 + 
\mid S_{{\bf p},-}(t) \mid^2 \right).   
\label{ioniz_prob_of_center_A}
\end{eqnarray}    
Similar expressions hold for ionization of center $B$. 
Note that the time integrals in (\ref{ioniz_of_center_A-add}) 
are easily taken analytically. However, the resulting 
expressions are somewhat lengthy and will be given elsewhere. 

Let us now turn to the discussion of some results following from 
the above expressions. Perhaps one of the most appropriate 
objects, where the process under consideration may occur, 
is a helium dimer exposed to a resonant field. However, 
for simplicity we restrict our attention here to  
the most fundamental two-atomic system consisting 
of two hydrogen atoms, which are initially 
in the ground state and are irradiated by an electromagnetic field 
with frequency resonant to the atomic $1s$-$2p$ transition. 
The internuclear distance $R$ is assumed to be such that 
the changes in the atomic levels, 
caused by the (full) interatomic interaction,  
remain smaller than even the natural width $\Gamma_r$ 
of the $2p$ level in atomic hydrogen, which is fulfilled 
at $R \stackrel{>}{\sim} 20$ a.u.. Since we also 
suppose that $\left| W_{01} \right| \gg \Gamma_r$ 
(which in case of hydrogen holds for 
$F_0 \stackrel{>}{\sim} 10^{-7}$ a.u.) 
these changes are negligible compared to those 
induced by the electromagnetic field. 
Therefore, one can indeed treat the ionization process by 
regarding the two-atomic system as consisting 
of two individual hydrogens whose atomic properties 
remain basically unchanged.   
\begin{figure}[b]  
\vspace{-0.25cm}
\begin{center}
\includegraphics[width=1.05\columnwidth]{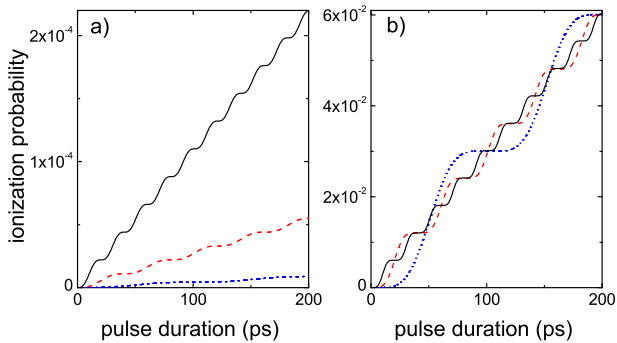} 
\end{center}
\vspace{-0.5cm} 
\caption{ \footnotesize{ Probability for single ionization 
of two hydrogen atoms separated by a distance $R=25$ a.u. 
given as a function of the electromagnetic pulse duration 
for zero detuning. The interatomic vector ${\bf R}$ is directed along 
the field polarization ${\bf F}_0$.   
Dot, dash and solid curves show the results for 
$F_0=2 \times 10^{-6}$, $5 \times 10^{-6}$ and 
$10^{-5}$ a.u., respectively. 
a) the contribution of the direct channel only. 
b) the contribution of the two-center channel. } } 
\label{figure2}
\end{figure}

For not too strong electromagnetic fields and 
not too large separations between the atoms 
the two-center channel can strongly outperform the direct channel 
in photo ionization. Indeed, neglecting the retardation 
effects, one can show that for ionization of center 
$A$ the relative strength of the former with respect 
to the latter is determined by the ratio $\left( d_B/(R^3 F_0) \right)^2 $, 
where $d_B$ is the dipole moment of the bound-bound 
transition in center $B$. 
The case, when this ratio substantially 
exceeds unity, is illustrated 
in figure \ref{figure2} where we present 
the probability $P_A(t) + P_B(t)$ for single ionization 
of two centers as a function of time. 
Since electron transitions into the continuum states 
occur only from the excited states 
the probability shows a non-monotonous behavior 
in which time intervals, 
when the ionization probability rapidly increases, 
are separated by intervals, when the probability 
remains practically constant, reflecting oscillations 
with the Rabi frequency of the electron populations between  
the ground and excited states in a resonant electromagnetic field. 

Such a staircase behavior, related to the Rabi oscillations, 
is inherent for both the direct \cite{knight-review}  
and two-center channels of ionization 
(see figure \ref{figure2}a and \ref{figure2}b, respectively).  
However, since in the latter case for ionization to occur 
both centers have to be in the excited states, 
the "stairs"  in the time development 
of the probability for ionization via ICD are more pronounced. 

Note also that the probability for ionization 
via ICD may demonstrate a behavior which appears 
counterintuitive at first glance: 
the weakest field can lead to the highest instantaneous 
value of this probability (see figure \ref{figure2}b, 
pulse durations in the intervals around $75$ and $175$ ps). 
However, a simple analysis shows that 
such a behavior is in fact the consequence 
of (i) the dependence of the Rabi frequency 
on the field intensity and 
(ii) the field-independence of 
the probability for ionization via the two-center channel,  
temporally averaged over the inverse of the Rabi frequency.   

\begin{figure}[b]  
\vspace{-0.25cm}
\begin{center}
\includegraphics[width=0.44\textwidth]{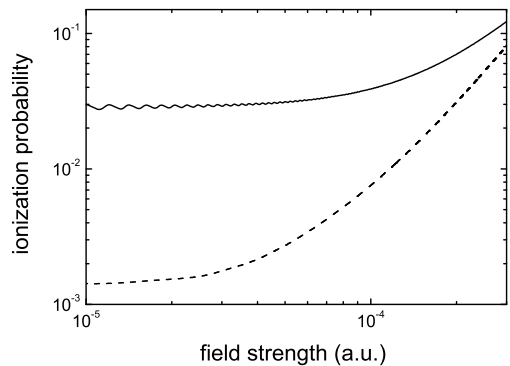} 
\end{center}
\vspace{-0.5cm} 
\caption{ \footnotesize{ Probability for single ionization 
of two hydrogen atoms separated by a distance $R=25$ a.u. given 
as a function of the electromagnetic field strength for a pulse duration of 
$100$ ps. ${\bf R}$ is directed along ${\bf F}_0$.   
Solid curve: ionization by a monochromatic field 
with zero detuning. Dash curve: ionization by 
a non-monochromatic field with bandwidth of $20$ meV 
and central frequency equal to the $1s$-$2p$ transition frequency 
in hydrogen.} } 
\label{figure3}
\end{figure}
 
The direct and two-center channels of photo ionization 
are in general characterized by different dependences 
on the field intensity. For instance, in the case of 
ionization by a monochromatic field, which is exactly resonant to 
the bound-bound transitions, 
the population probabilities for the excited bound states  
(averaged over the inverse of the Rabi frequency)  
do not depend on the field intensity. 
As a result, the averaged probability for photo ionization 
via the direct channel, 
which involves absorption of an additional photon, 
is proportional to the field intensity. 
In contrast, the averaged probability for ionization due to 
the two-center mechanism is intensity-independent because 
it relies only on resonant bound-bound transitions 
and two-center electron-electron correlations.  
These two limiting dependences and a smooth transition between 
them are illustrated by the solid curve in figure \ref{figure3}.    
Similar results hold also for non-monochromatic 
fields (see dash curve in figure \ref{figure3}).  
At higher intensities, where the direct channel 
is more efficient, the ionization probability demonstrates 
a linear growth with intensity 
whereas at smaller intensities, 
where the two-center mechanism dominates, 
the ionization probability becomes almost a constant. 
The wiggles seen in figure \ref{figure3} result from 
the Rabi-flopping dynamics. 

\begin{figure}[b]  
\vspace{-0.25cm}
\begin{center}
\includegraphics[width=1.075\columnwidth]{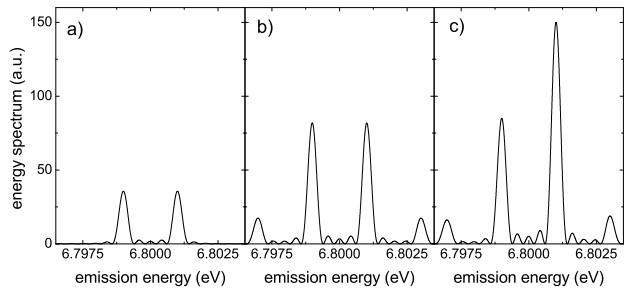}
\end{center}
\vspace{-0.5cm} 
\caption{ \footnotesize{ Energy spectrum of electrons emitted 
in the process of single ionization of two hydrogen atoms 
separated by a distance $R=25$ a.u. at $F_0 = 10^{-4}$ a.u., 
zero detuning, pulse duration of $10$ ps 
and ${\bf R}$ directed along ${\bf F}_0$.  
Panels a) and b) show the partial contributions 
of the direct and two-center mechanisms, respectively. 
Panel c) displays the total spectrum 
including the interference of the two channels. }} 
\label{figure4}
\end{figure}

In a resonant electromagnetic field each 
of the ground and excited levels of 
the centers $A$ and $B$ split into two sub-levels. 
Therefore, the direct ionization channel 
leads in general to two Autler-Townes lines 
(per atomic species) in the energy spectrum of 
the emitted electrons located at 
$ \varepsilon_p = z_{\pm}^A + 2 \omega_0$ 
(see figure \ref{figure4}a).  
Because of the splitting electron 
transitions between dressed bound states 
are characterized by three different transition energies. 
When these energies are transferred 
to the neighbor field-dressed center via ICD, 
one would obtain in general six energy lines (per atomic species) 
for the emitted electrons. For the emission from center $A$ they are given by  
$ \varepsilon_p = z_{\pm}^A + 2 \omega_0$,  
$ \varepsilon_p = z_{\pm}^A + z_{-}^B - z_{+}^B + 2 \omega_0$ 
and 
$ \varepsilon_p = z_{\pm}^A + z_{+}^B - z_{-}^B + 2 \omega_0$.   
However, for identical centers only four of 
them have different energies (see figure \ref{figure4}b).   

Since in the transition amplitude $S_{{\bf p},+}$ 
both direct and ICD ionization channels 
may lead to the same final states, they can interfere. 
As was mentioned, for identical atoms the direct and 
ICD channels result in two and four emission lines, respectively, in the energy spectrum 
of the emitted electrons. The former two coincide with the two central 
lines from the latter four. Therefore, for these two lines 
interference may occur. Indeed, comparing 
figures \ref{figure4}a, \ref{figure4}b and \ref{figure4}c 
we see that the total contribution of the two channels 
is not simply equal to the sum of their partial contributions. 
In particular, destructive and constructive interferences 
of these contributions in the amplitude $S_{{\bf p},+}$ 
occur, respectively, for the left and right central peaks.    

In conclusion, resonant two-photon single ionization of 
a two-atom system was considered. For simplicity and 
in order to clearly reveal the basic underlying physics 
results for two hydrogen atoms were shown. 
It was found that, in a certain range of interatomic distances and
external field strengths, the ionization is dominated by the two-center
channel involving ICD. In this case, the mean ionization yield - averaged over
one Rabi oscillation - becomes independent of the applied field intensity. The
instantaneous ionization probability shows a step-wise increase with
time and a non-monotonous dependence on the field strength. The
photo-electron spectrum comprises four lines. Its asymmetry reflects the
interference between the direct and two-center ionization channels.
An experimental observation of the predicted effects 
can be possible utilizing helium dimers, which were recently 
used successfully for studies of various aspects of ICD \cite{helium-dimers}.

\end{document}